\documentclass[12pt]{iopart}
\usepackage[utf8]{inputenc}
\usepackage{iopams}  
\usepackage[pdftex]{graphicx}
\usepackage{dcolumn}

\begin{document}

\title{Effect of edge dislocations on structural and electric properties of 4H-SiC} 

\author{J.~{\L}a\.{z}ewski, P. T. Jochym, P. Piekarz, M. Sternik, K.~Parlinski,}
\address{Institute of Nuclear Physics, Polish Academy of Sciences,
             Radzikowskiego 152, 31-342 Cracow, Poland}
\author{J. Cholewi\'{n}ski, P. D{\l}u\.{z}ewski}
\address{Institute of Fundamental Technological Research, 
            Polish Academy of Sciences, Pawi\'{n}skiego 5b, 02-106 Warsaw, Poland}
\author{S. Krukowski}
\address{Institute of High Pressure Physics, Polish Academy of Sciences,  
            Soko\l{}owska 29/37, 01-142 Warsaw, Poland}
\address{Interdisciplinary Centre for Materials Modelling, 
            Warsaw University, Pawi\'{n}skiego 5a, 02-106 Warsaw, Poland}

\ead{jan.lazewski@ifj.edu.pl}
\vspace{10pt}

\begin{abstract}
The paper presents a study of two full-core, edge dislocations of opposite Burgers vectors in 4H-SiC, conducted using the first-principles density functional theory methods.
We have determined the creation energy of the dislocations as a function of distance between their cores. The radial distribution function has been applied to examine strong impact of the dislocations on the local crystal structure. 
The analysis of the electronic structure reveals mid-gap levels induced by broken atomic bonds in the dislocation core. The maps of charge distribution and electrostatic potential have been calculated and the significant decrease of the electrostatic barriers in the vicinity of the dislocation cores has been quantified.
The obtained results have been discussed in the light of available experimental data. 
\end{abstract}


\maketitle

\section*{Introduction}

The Silicon Carbide (SiC), a wide band gap semiconductor, is a promising material for high-voltage and high-frequency nanoelectronic devices \cite{zopler2005,choyke2004}. 
Very good operational quality of SiC results from high values of breakdown voltage ($\approx10^6$ V/cm), high charge carrier mobility, high temperature stability and high thermal conductivity \cite{bechstedt1995}.
Additionally, this material has very good mechanical properties and resistance to radiation damage. 
Unfortunately, the electronic properties of epitaxial layers strongly depend on the material quality. 
The presence of intrinsic defects and impurities which arise during crystal growth process substantially limit applications of SiC.

Dislocations are the main crystal defects in SiC. They deteriorate the performance of high electric field devices such as Shottky and $p$-$n$ diodes \cite{neudeck2000,singh2006}. Two typical components of dislocations with the direction along $[0\,0\,0\,1]$ are the screw and edge dislocations with Burgers vectors $[0\,0\,0\,1]$ and $\frac{1}{3}[\bar{2}\,1\,1\,0]$, respectively. The standard density of dislocations observed in the good quality 4H-SiC epilayers reaches $10^{3}$ cm$^{-2}$ \cite{chen2008}. The electrical characteristics of 4H-SiC photodiodes reveal that the screw and edge dislocations reduce the breakdown voltage by $3.5\%$ and $2\%$, respectively, and increase the leakage currents compared to systems without crystal defects \cite{berechman2010}. Dislocations also influence transport properties of 4H-SiC by increasing recombination activity \cite{maximenko2004} and reducing the diffusion length in the material \cite{maximenko2010}.

Various mechanisms are responsible for modifications of electronic structure in the presence of dislocations. 
In the dislocation core, the broken bonds give rise to acceptor levels within the band gap and make this region electrically active \cite{shockley1953,pearson1954}. 
Dislocations also act as trapping centres and sources for point defects \cite{schroter1995}. 
In the vicinity of the core, the strain field may induce additional states located near the conduction band edge \cite{celli1962}. 
In epitaxial SiC, several gap levels have been detected using the deep-level transient spectroscopy (DLTS) and their connection with the intrinsic defects have been analysed \cite{danno2007,sasaki2011}. 

The structural properties of dislocations have been previously studied theoretically in very few materials using both empirical potentials \cite{duesbery1991} and {\it ab initio} methods based on the density functional theory (DFT) \cite{bigger1992,arias1994,blase2000,cai2001}. 
The DFT provides tools that allow one to study changes in electronic structure induced by dislocations \cite{kontsevoi2001}. 
Such computations revealed that the dislocation core in GaN induces deep-gap states \cite{lee2000,lymperakis2004}, which strongly influence the electronic and optical properties of this semiconductor \cite{you2007}.
The point defects such as vacancies and oxygen dopants can be easily trapped at the core of the edge dislocations providing additional mid-gap states \cite{lee2000,elsner1998}. 
The influence of partial dislocations on the structural and electronic properties have been investigated in the pure SiC \cite{blumenau2003} and crystals doped with impurity atoms \cite{bernardini2005}.
However, to our best knowledge, electronic properties of the full-edge dislocation has not been studied with {\it ab initio} methods yet.

In this work we study the changes in the crystal structure and electronic properties of 4H-SiC induced by the edge dislocation. 
We analyse in detail the local lattice distortion by means of the radial distribution function (RDF) -- the density of atoms in the spherical shell around given atom averaged over whole structure \cite{chandler1987,hansen2013,kittel1986}.
The calculations disclose significant redistribution of charges and electrostatic potential in the region of the dislocation core.
The electronic states located in the insulating gap, arising from the dislocation core atoms, show a very weak dispersion in the perfect correlation with the distribution of charges in the distorted region.
Estimated minimal electrostatic barriers for ideal and distorted systems throw a new light on the decrease of the breakdown voltage in the defected 4H-SiC crystal. 

\section*{Methods}

All calculations have been executed with the Vienna Ab-Initio Simulation Package (VASP) \cite{kresse1996a,kresse1996b} using the full-potential projector-augmented wave method \cite{blochl1994,kresse1999} within the generalised-gradient approximation (GGA) \cite{perdew1992,perdew1993,perdew1996}. 
The following valence base configurations have been included: Si $3s^23p^2$ and C $2s^22p^2$. The integration over the {\bf k}-point space has been performed over the $2\times2\times2$ Monkhorst-Pack mesh \cite{monkhorst1976} and the energy cut-off for the plane waves expansion was equal to 500 eV. 
The structures of ideal and defected crystals have been fully optimised (with respect to lattice parameters, stresses, and internal degrees of freedom) in the supercells consisting of 8$\times$6$\times$1 primitive unit cells, containing 384 and 346 atoms, which correspond to 25\AA$\times$19\AA$\times$10\AA{} and 22\AA$\times$19\AA$\times$10\AA{} crystal volumes, respectively. 
We have used the conjugate gradient technique with the energy convergence criteria set at $10^{-8}$ eV and $10^{-5}$ eV for the electronic and ionic iterations, respectively.

\section*{Results}
\subsection*{Crystal structure}

%
\begin{figure}
\begin{center}
\includegraphics[width=0.5\columnwidth]{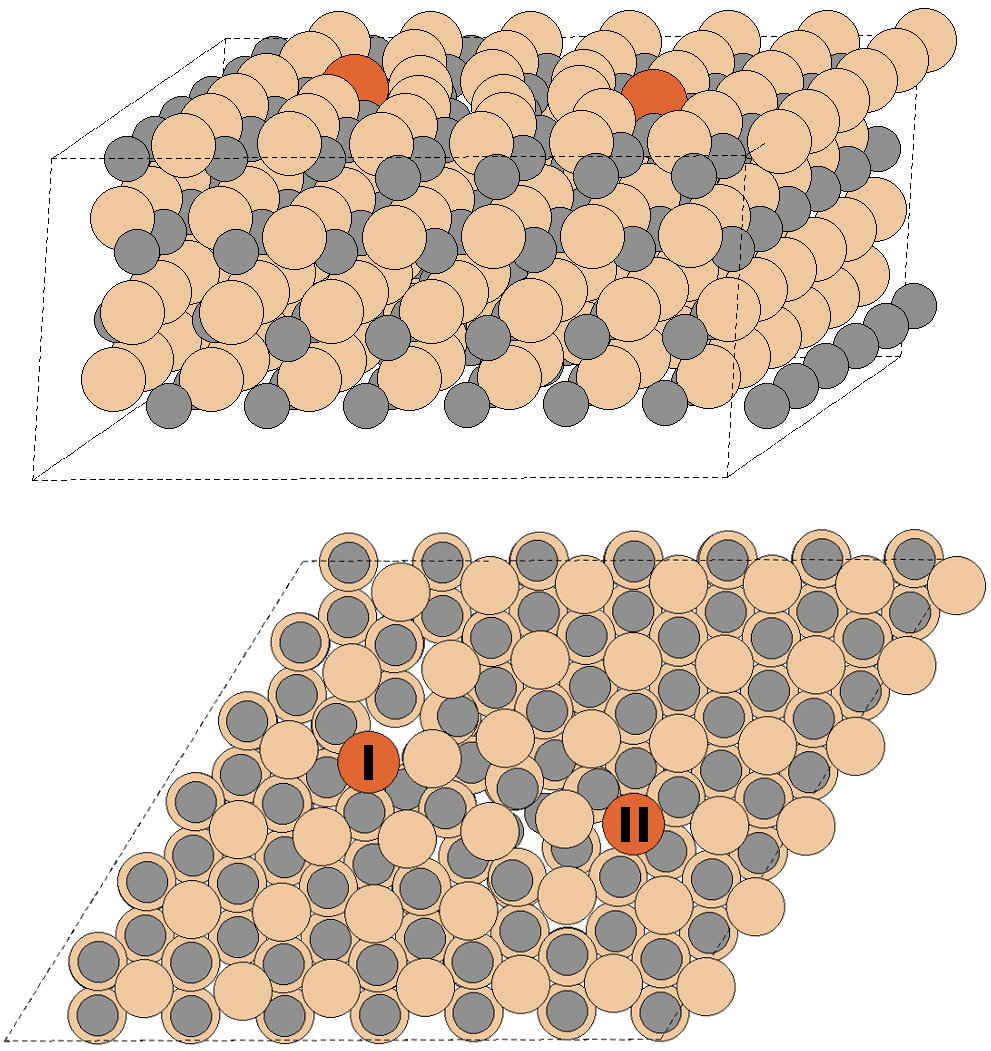}
\end{center}
\caption[]{\label{fig:structure}Off-bottom (upper panel) and top (lower panel) view of the optimised crystal structure of 4H-SiC with two full-core edge dislocations. Both dislocation edges (I and II) have been distinguished by colour.}
\end{figure}
The pure 4H-SiC crystal structure has been reproduced correctly with the lattice constants $a=3.095$ \AA{} and $c=10.132$ \AA, which are very close to experimental values that amount to 3.073 \AA{} and 10.053 \AA, respectively.
To fulfil the periodic boundary conditions we consider two full-core edge dislocations with opposite Burgers vectors $\frac{1}{3}[\bar{2}\,1\,1\,0]$ and $\frac{1}{3}[2\,\bar{1}\,\bar{1}\,0]$, inserted into the system using Visual Editor of Crystal Defects (VECDs) \cite{vecds}.
It involves removal of 38 selected atoms (19 SiC molecules) changing the total number of them, which has been taken into account in all comparisons to the ideal structure. 
Nevertheless, the relaxed crystal volume noticeably increases (by about 1\%) due to the local destruction of hexagonal close-packed structure. 
However, since the supercell is big enough, the elongation of the average near neighbour bond is below 0.5\% in comparison to its length in the ideal system. 

The creation energy per unit length of a single edge dislocation has been calculated as a function of distance between their cores using the following formula \cite{lymperakis2004}:
\[e=(E-E'_0)/2c,\] where $E$ is the energy of the system with a pair of dislocations (346 atoms), $E'_0$ is the ground state energy of the undisturbed system normalised to the same number of formula units. For relative distance between cores equal to 6.73 \AA{} we have obtained $e=2.17$ eV/\AA{} per supercell. Additionally, it has been found that $e$ increases for larger distances due to attractive force between two dislocations of opposite sign \cite{hirth1982}. The Peierls energy barrier between two states of dislocation core at adjacent lattice nodes along the common glide plane is very low. Therefore, to prevent the annihilation of defects during optimisation process we have trapped both dislocations by pinning their cores at different glide planes.

The optimised 4H-SiC structure with two edge dislocations is presented in Fig.~\ref{fig:structure}. The positions of dislocation lines have been marked by atoms distinguished by different colour. 
Carrying out crystal structure relaxation, without any symmetry constrains, we have found that it tends to self-recover -- reconstructing as much as possible of pure 4H-SiC coordinations and so gathering most of the disorder to one of the dislocation cores (I in Fig.~\ref{fig:structure}).

%
%
\begin{figure}
\begin{center}
\includegraphics[width=0.7\columnwidth]{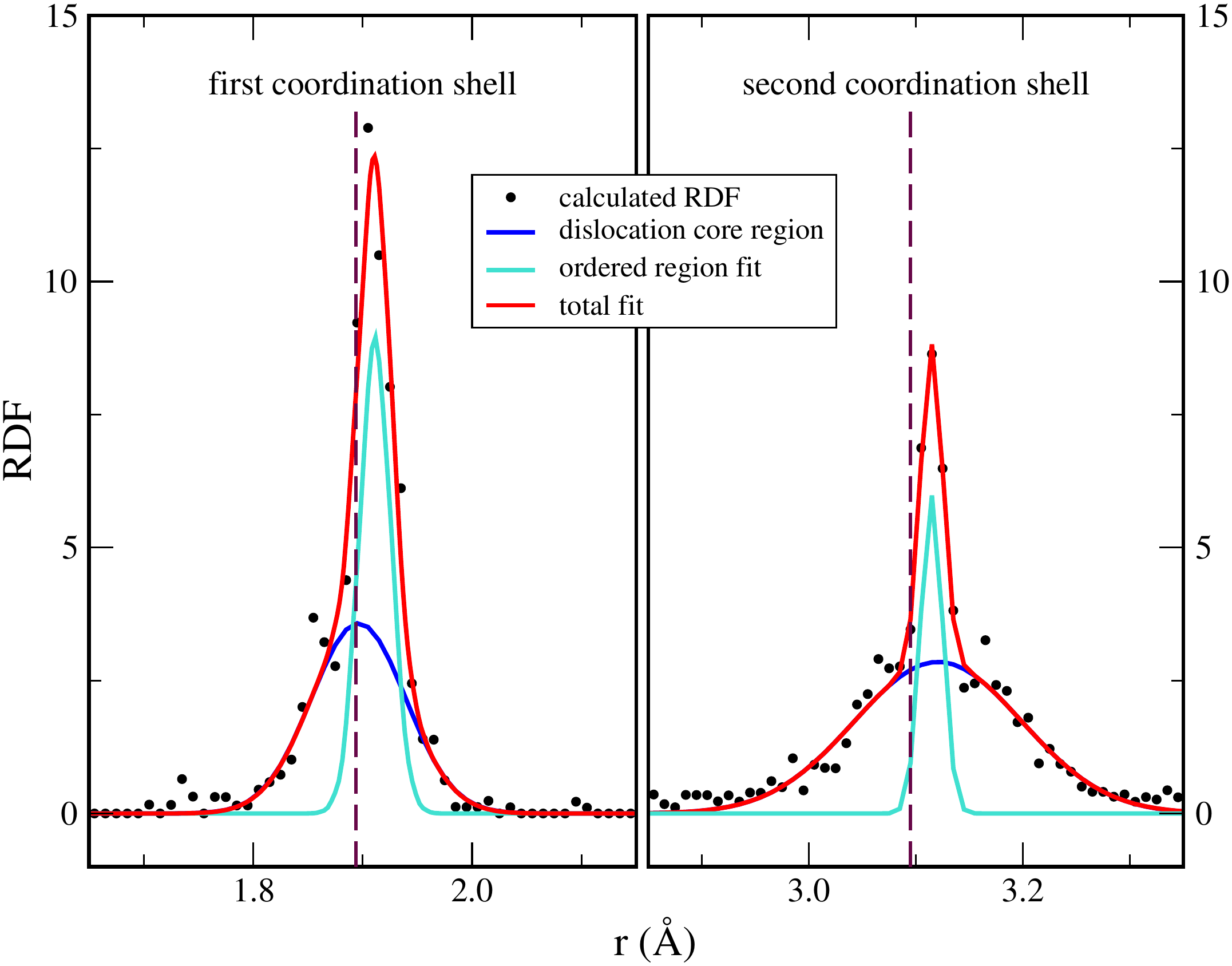}
\end{center}
\caption[]{\label{fig:rdf}RDF of the first-(left panel) and the second-coordination shell (right panel) for the edge dislocation dipole in 4H-SiC (points), fitted with the two-component Gaussian (solid lines).
The dashed lines denote positions of the coordination shells for the ideal 4H-SiC system.}
\end{figure}

To analyse the influence of the dislocation on its nearest neighbourhood, the RDF for the dislocation dipole has been calculated and analysed. 
In Fig.\ \ref{fig:rdf} one can immediately notice that the introduction of the edge dislocation disturbs hexagonal close-packed structure which results in slight increase of the near neighbour and next near neighbour distances visible as shift from pure 4H-SiC positions (dashed lines). 
We discovered that each coordination shell can be fitted with a two-component function (two Gaussian lines with parameters collected in Table~\ref{tab:rdf}): the central part with small width and the rest reproducing broad tails behaviour. 
In particular this is clearly visible for the second coordination shell (right panel in Fig.~\ref{fig:rdf}). 
%
\begin{table}
\caption{\label{tab:rdf}The parameters of Gaussian functions (positions and FWHM) and calculated peak areas of two-component fit of RDF presented in Fig.~\ref{fig:rdf}. u and d parts corresponds to the undistorted and distorted regions, respectively.}
\begin{indented}
\item[]\begin{tabular}{@{}l c c c c}
\br
         & \multicolumn{2}{c}{near neighbour}        & \multicolumn{2}{c}{next near neighbour}    \cr
           \cline{2-5}
         & u           & d               & u          & d           \\\ns
\mr
position & 1.911       & 1.896           & 3.115      & 3.121         \\
area     & 0.321       & 0.387           & 0.154      & 0.560         \\
FWHM     & 0.034       & 0.101           & 0.024      & 0.185         \\
\br
\end{tabular}
\end{indented}
\end{table}

We have found that the components with a narrow and broad full width at half maximum (FWHM) correspond to the undistorted part of the crystal and disordered dislocation core, respectively.
For bigger supercells with smaller densities of dislocations the latter component will remain mostly unchanged while the former one will increase and converge towards a delta function.

\subsection*{Band structure}

%
\begin{figure}[t]
\begin{center}
\includegraphics[width=0.75\columnwidth]{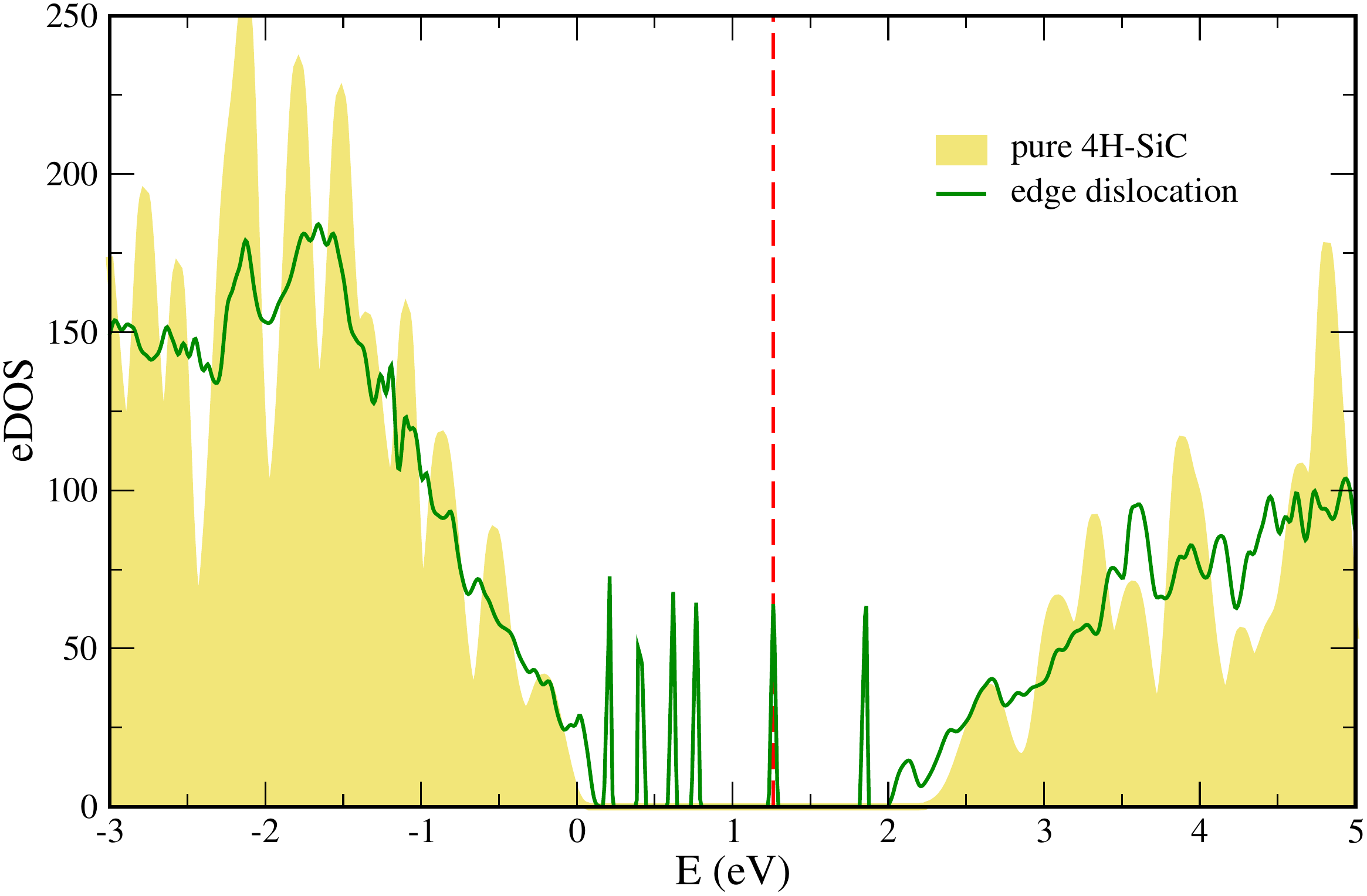}
\end{center}
\caption[]{\label{fig:bands}The electronic density of states (eDOS) calculated for pure 4H-SiC (filled contours) compared with results obtained for the distorted system (thin colour lines). Fermi energy level has been marked with dashed line.}
\end{figure}

Next, we analyse the influence of edge dislocations on the band structure of the 4H-SiC. 
The general observation is that the positions of the main bands in the electron density of states (eDOS) remain mostly unchanged (see Fig.~\ref{fig:bands}). 
The characteristic peaks in the valence and conduction bands of the ideal structure become broadened and smeared out in the system with dislocations. 
The strongest changes are found inside the energy gap of the pure 4H-SiC crystal\footnote[1]{The theoretical gap $E_g=2.3$ eV is smaller than the measured value (3.2 eV) due to well-known gap underestimation problem in the LDA/GGA approaches.}, where additional localised states appear. 
One can find sharp deep states delivered by atoms with broken nearest neighbours bonds and broadened shallow bands close to the top of the valence band and the bottom of the conduction band. 
The presence of shallow bands effectively reduces the magnitude of the insulating gap.
The states located in the range $\sim0.2-0.6$ eV below the conduction band have also been found in the SiC systems with the stacking faults \cite{miao2001,iwata2001} and Shockley partial dislocations \cite{blumenau2003}. The latter induce also the occupied states about 0.4 eV above the valence band. All deep states are very narrow which indicates their weak dispersion in the Brillouin zone and localised character. 
It is possible to trace the distribution of individual states in the whole spectrum, as well as to identify atoms which give their contribution to the eDOS in a particular energy range. 
We have found that the deep states originate from the atoms in the distorted region of the dislocation core. 
The detailed analysis has led to the conclusion that also partial eDOS projected to the $spd$ states has been modified. For the conduction band, where in the perfect 4H-SiC $d$ states dominate, as well as for the localised states inside the gap, the main contribution for distorted system stems from $p$ states.

\begin{figure}
\begin{center}
\includegraphics[width=0.8\columnwidth]{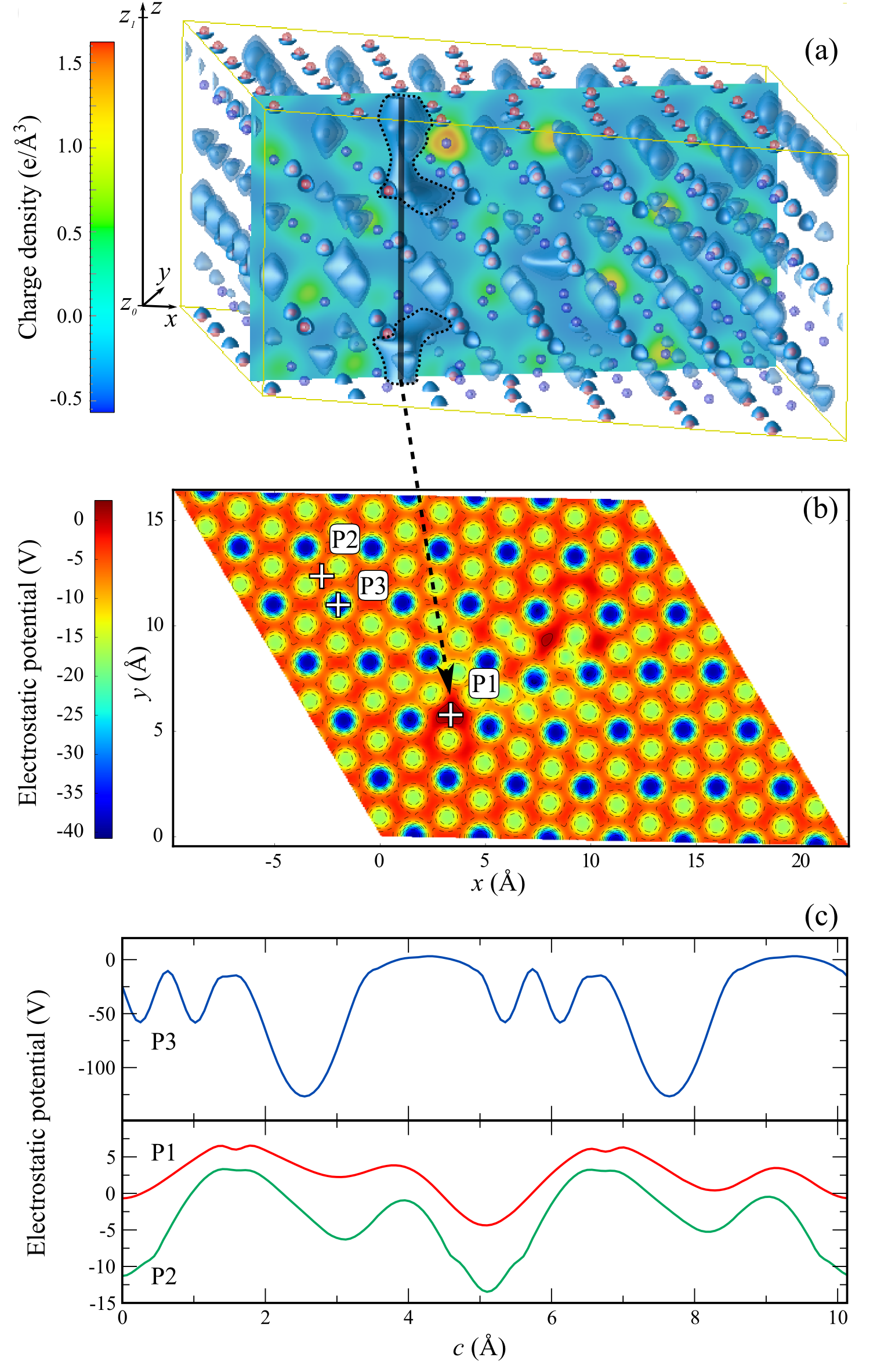}
\end{center}
\caption[]{\label{fig:charges}(a) Off-bottom view of charge distribution in distorted system. The slice plane crosses through both edge dislocations cores. 
(b) Top view of electrostatic potential derived from charge distribution averaged along $c$ direction. 
(c) The electrostatic potential along $c$-axis direction in P1 (dislocation core), P2 (between atoms in the ordered region), and P3 (atomic chain) places marked in panel (b).}
\end{figure}

\subsection*{Electric charge distribution}

The charge density is a well-defined, primary quantity in the DFT. Fig.~\ref{fig:charges}a presents the charge redistribution around edge dislocations. 
In particular, one can see the elongated structures with a very low charge density (outlined with dotted line in Fig.~\ref{fig:charges}a) in the slice plane crossing both dislocation cores. 
On the other hand, the neighbouring atoms visible in Fig.~\ref{fig:charges}a have increased density of carriers connected with broken interatomic bonds. 
The effect is more pronounced for the dislocation I, marked with the thick solid line in the figure.
The electrostatic potential derived from the charge distribution and averaged over $c$ direction is presented in the top view projection (Fig.~\ref{fig:charges}b). 
The decreased charge density regions exhibit lowered energy barriers which allow for easier flow of the carriers. 
Again, the effect is much stronger for the dislocation I with the bigger disorder around its edge.
Additionally, in panel (c) the electrostatic potential along the $c$ direction in three different locations of distorted system has been shown. Its periodicity reflects ABCB stacking in the 4H-SiC structure. In accordance with the intuition, electric charges encounter the highest barriers along the atomic chain (curve P3 in Fig.~\ref{fig:charges}c) and the lowest through the dislocation core (curve P1). 

To quantify the influence of dislocations on the energy barriers for the electron transport along $c$-axis, we have analysed the local electrostatic potential in the crystal. 
We have taken into account that the path over minimal barriers can wander not necessarily along a straight line parallel to the $c$ direction.
Therefore, for each pair of positions: $r_0=(x,y,z_0)$ and $r_1=(x,y,z_1)$ -- where $z_0$ is at the bottom of the periodic unit while $z_1$ is at its top (see Fig.~\ref{fig:charges}a) -- we have derived a minimal energy path and the barrier height for this path from $r_0$ to $r_1$. 
The calculation used the basin-filling segmentation algorithm \cite{pal1993} with bisection search for minimal limiting energy level.
The results indicate substantial lowering of the barrier in the vicinity of the dislocation cores denoted by P1 in Fig.~\ref{fig:charges}b. 
The calculated difference between the minimal barrier height in the non-distorted area and in the vicinity of the dislocation core is 0.8~V. 
This value is much lower than the differences in barrier heights presented in Fig.~\ref{fig:charges}c due to the nonlinear character of the minimal barrier path. 
The calculated energy barriers are obtained with zero voltage bias and as such are not sufficient to determine quantitatively the charge transport properties of the material but are an important step towards full understanding of charge carriers' behaviour.

On the basis of the presented results, the following mechanism of the insulating properties weakening and the breakdown voltage decreasing in the 4H-SiC monocrystal may be proposed. Very low energy barriers along glide planes cannot oppose the formation of dislocations around point defects and local lattice stresses as well as their migration through the crystal. In a dislocation core, part of interatomic bonds are broken causing the creation of deep states inside the semiconductor gap. Additionally, crystal distortions modify the atomic potential enforcing the shallow states formation and consequently narrowing the forbidden gap. Finally, the elongated regions of the reduced charge density with flat electrostatic potential are formed along a dislocation core. The lack of barriers inside so created tunnels, in connection with additional states in the energy gap, may enhance a carrier flow through the distorted areas, significantly influencing decrease in the breakdown voltage. This scenario provides the plausible explanation of the current-voltage characteristics of the 4H-SiC avalanche photodiodes, which show a pronounced decrease of the breakdown voltage and increase of leakage currents due to a single edge dislocation \cite{berechman2010}. Our analysis of the mid-gap levels also enables a better understanding of a pronounced impact of dislocations on the carrier lifetime \cite{maximenko2004} and diffusion length \cite{maximenko2010}, and can be helpful in interpreting the DLTS measurements \cite{danno2007,sasaki2011}. 
Moreover, similarly to other strongly defected systems \cite{wdowik2013}, the changes in the electronic structure may be related to observable effects in the lattice dynamical and optical properties of SiC \cite{talwar2015}.

\section*{Conclusions}

Summarising, in this work we have successfully modelled a pair of edge dislocations using {\it ab initio} methods. 
We have confirmed experimental findings concerning small but well-visible lattice constants elongation, energy gap narrowing, and electrostatic barriers reduction. 
Furthermore, with atomic-scale resolution, we have explained foregoing processes. 
We have shown that (i) the crystal structure is strongly disturbed in the small vicinity of the dislocation core, (ii) additional energy levels occurring in the energy gap belong to the atoms with broken bonds occupying the core neighbourhood, (iii) existence of spatial tunnels, with atoms delivering localised states to the band structure on its sides, significantly decreases electrostatic barriers and should be considered as one of the primary factors responsible for experimentally observed reduction of breakdown voltage.

\ack
This work was partially supported by the SICMAT Project financed under the European Founds for Regional Development (Contract No. UDA-POIG.01.03.01-14-155/09).

\section*{References}
\bibliography{SiC-dislo}

\end{document}